\documentstyle[twocolumn,amssymb,aps,psfig,float]{revtex}
\begin{document}
\bibliographystyle{plain}
\twocolumn[\hsize\textwidth\columnwidth\hsize\csname @twocolumnfalse\endcsname

\title{ 
Low-energy sector of the S=1/2 Kagome antiferromagnet
}
\vskip0.5truecm 
\author{F. Mila}

\address{Laboratoire de Physique 
Quantique, Universit\'e Paul Sabatier, 118 Route de Narbonne, 31062 
Toulouse Cedex, France. }\vskip0.5truecm
      
\maketitle

\begin{abstract}
\begin{center} 
\parbox{14cm}
{Starting from a modified version of the the S=1/2 Kagome 
antiferromagnet to emphasize the role of elementary triangles, an effective
Hamiltonian involving spin and chirality variables is derived. A mean-field
decoupling that retains the quantum nature of these variables is shown to yield
a Hamiltonian that can be solved exactly, leading to the following predictions:
i) The number of low lying singlet states increase with the number of sites $N$
like $1.15^N$; ii) A singlet-triplet gap remains in the thermodynamic limit;
iii) Spinons form boundstates with a small binding energy. 
By comparing these properties with those of the regular Kagome lattice as
revealed by numerical experiments, we argue that this description captures the
essential low energy physics of that model.
}
\end{center}
\end{abstract}
\vskip .1truein
 
\vskip2pc
]

Despite a very intense activity over the past 10 years, the magnetic properties
of the S=1/2 Kagome antiferromagnet remain an open problem. If a number of facts
seem to be rather firmly established by now thanks to the very extensive
numerical simulations that have been performed on that
system\cite{zeng,chalker,leung,singh,elstner,nakamura,lecheminant,bernu}, 
a simple 
theoretical picture that accounts for the basic findings has not emerged yet.
The most striking feature is probably the presence of many, low--lying singlet
states\cite{bernu}. The first indication that this might be the case was the 
appearance of a
low temperature peak in the specific heat. While the evolution of this peak
with the size of the system is not clear yet, the numerical
determination of all the low-lying singlet states for systems with up to 36
sites shows that their number increases like $1.15^N$, where $N$ is the number
of sites of the system. The
best candidate to explain this proliferation of low-lying singlets 
is a short-range RVB
description of the low-energy sector based on dimer
coverings of the Kagome lattice with nearest-neighbour singlets
\cite{elser,sachdev,zeng2}. The main
problem with this approach is that the number of dimer states increases 
like $1.26^N$, i.e. much too fast\cite{note1}, and no convincing criterion 
could be 
found that allows one to select the relevant singlet states. 
The other
important, although less accurately established, findings of the numerical 
simulations are the absence of long-range magnetic order in the ground-state and
the presence of a singlet--triplet gap in the thermodynamic limit\cite{bernu}. 
Finally the 
role of spin 1/2 excitations, as well as the consistency of the numerical
results with some exotic types of order\cite{kalmeyer,marston,yang,chandra}, 
is still under investigation.

In this paper, we propose a simple explanation of these properties. We start 
from
the following observation:
The exponential increase of the number of these 
low--lying 
states suggest that they originate from the partial lifting of a local 
degeneracy that would be present if some of the couplings were set to zero. Now
the natural bricks to construct the Kagome lattice are triangles, and spins 1/2
on a triangle lead to a fourfold degenerate groundstate: two doublets that 
differ by their chirality. So let us investigate
how this degeneracy is lifted if one constructs the Kagome lattice by coupling
triangles. This amounts to studying the modified Kagome lattice
depicted in
figure 1 starting from the limit $J'/J\ll 1$. This can be seen as a triangular
lattice of triangles with $N_t=N/3$ sites, where $N$ is the number of sites of
the Kagome lattice.

\begin{figure}[hp]
\centerline{\psfig{figure=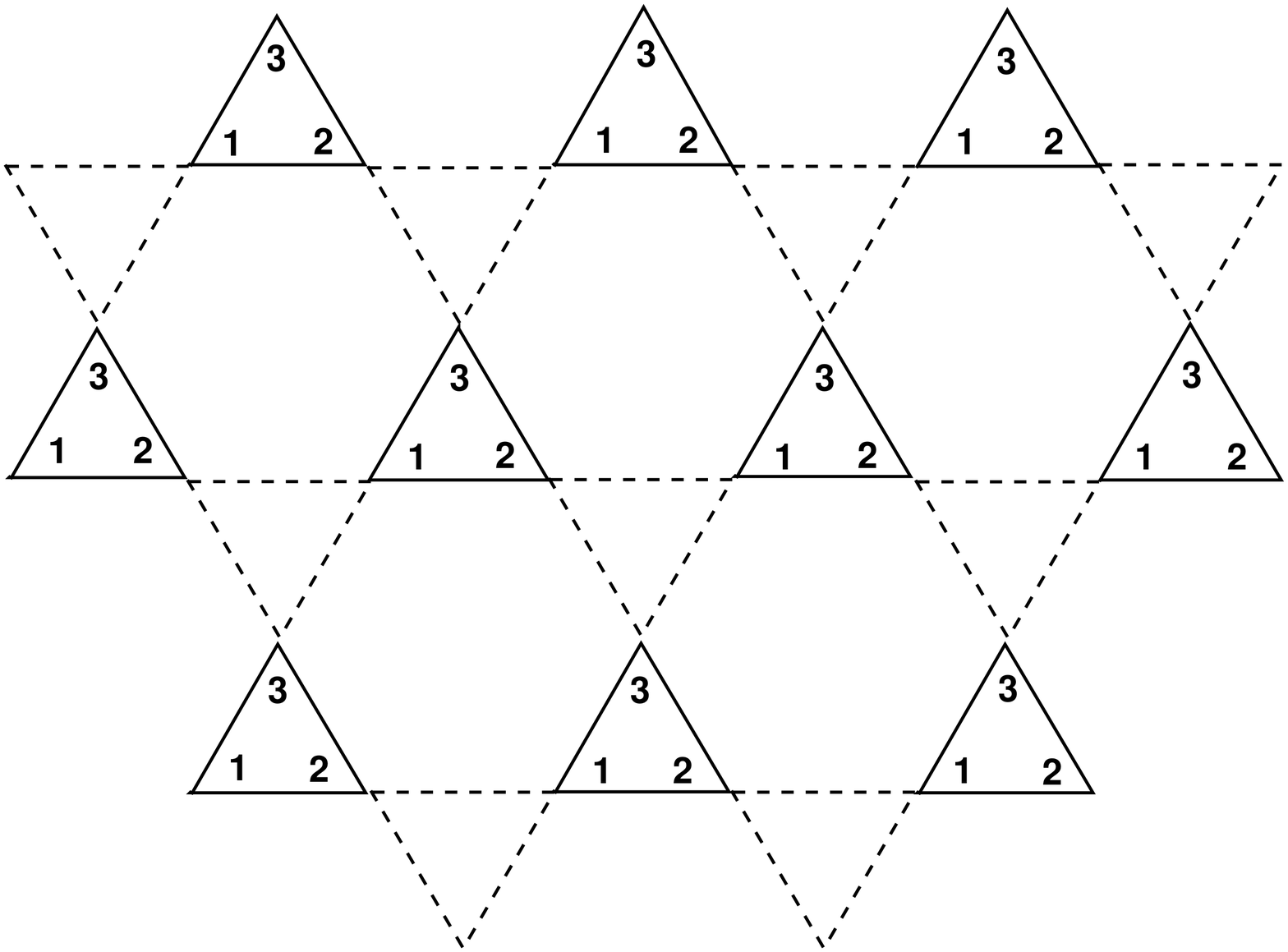,width=6.0cm,angle=0}}
\vspace{0.5cm}
\caption{Sketch of the dimerized Kagome lattice. Solid lines: $J$,
dashed lines: $J'$.}
\label{fig1}
\end{figure}

The first step is to derive an effective Hamiltonian in the subspace of the
groundstates of the triangles, as in Subrahmanyam's block spin
perturbation approach to the non-dimerized Kagome lattice\cite{subrahmanyam}.
Following
Schulz's approach to the 
problem of three coupled Heisenberg chains with periodic boundary
conditions\cite{schulz}, we describe the four groundstates of a triangle with 
two 
Pauli matrices: 
$\vec \sigma$ for
the spin of the doublet, the eigenstates of $\sigma_z$ being denoted
$\uparrow$ and $\downarrow$ , and $\vec \tau$ for its
chirality, the eigenstates of $\tau_z$ being denoted $R$ for right and $L$ for
left. 
In terms of
the original spins $\vec S$, these states can be written
\begin{eqnarray}
|\alpha R>=\frac {1} {\sqrt{3}} (|-\alpha \alpha \alpha > 
+\omega |\alpha -\alpha  \alpha >
+\omega^2 |\alpha \alpha -\alpha  >)\nonumber \\
|\alpha L>=\frac {1} {\sqrt{3}} (|-\alpha \alpha \alpha > 
+\omega^2 |\alpha -\alpha  \alpha >
+\omega |\alpha \alpha -\alpha  >)
\label{states}
\end{eqnarray}
where $\omega=\exp(2\pi i/3)$ and $\alpha = \uparrow$ or $\downarrow$. 
$|\alpha_1 \alpha_2 \alpha_3 >$ represents a configuration
of the original spins $\vec S$ within one triangle, the indices corresponding to
the convention of Fig. 1. 
Note that the total spin is now given by $(\sum'_i\vec \sigma_i)^2$, 
where the prime means that the sum runs over the triangular lattice. 
Each triangle has an energy $-3J/4$, and energies will be measured with 
respect to the
groundstate energy $-(3J/4)N_t$ of the $J'/J=0$ case.
Then, to
first order in $J'$, the effective Hamiltonian $\tilde H$ on the triangular
lattice is given by:
\begin{eqnarray}
\tilde H & = & (J'/9) \sum_{<i,j>} \tilde H^\sigma_{ij} \tilde H^\tau_{ij},\ \ 
\tilde H^\sigma_{ij}  = \sum_{<i,j>} \vec \sigma_i.\vec \sigma_j,
\nonumber \\
\tilde H^\tau_{ij} & = & 
(1-2(\alpha_{ij}\tau_i^- +
\alpha_{ij}^2\tau_i^+))(1-2(\beta_{ij}\tau_j^- + \beta_{ij}^2\tau_j^+))
\label{effective}
\end{eqnarray}
where $<i,j>$ denotes pairs of nearest neighbors.
In $\tilde H^\tau_{ij}$, $\alpha_{ij}$ and $\beta_{ij}$ are complex parameters
that depend on the type of bond:
$\alpha_{ij}$ (resp. $\beta_{ij}$) equals $1$, $\omega^2$ 
or $\omega$
when the original spin in triangle $i$ (resp. $j$) involved in
the bond $(i,j)$ sits at site
1, 2 or 3 with the convention of Fig. 1. 
In the basis $|RR>$, $|RL>$, $|LR>$ and $|LL>$ the eigenstates of 
$\tilde H^\tau_{ij}$ can be easily calculated:
\begin{eqnarray}
|\phi^\tau_{1}(i,j)>  =  \frac {1}{2} (1,-\beta_{ij},-\alpha_{ij},
\alpha_{ij} \beta_{ij} )\ \ \ \ E_1  =  9 \nonumber \\
|\phi^\tau_{2}(i,j)>  =  \frac {1}{2} (1,\beta_{ij},\alpha_{ij},
\alpha_{ij} \beta_{ij} )\ \ \ \ E_2  =  1 \nonumber \\
|\phi^\tau_{3}(i,j)>  =  \frac {1}{2} (1,-\beta_{ij},\alpha_{ij},
-\alpha_{ij} \beta_{ij} )\ \ \ \  E_3  =  -3 \nonumber \\
|\phi^\tau_{4}(i,j)>  =  \frac {1}{2} (1,\beta_{ij},-\alpha_{ij},
-\alpha_{ij} \beta_{ij} )\ \ \ \  E_4  =  -3
\label{taustates}
\end{eqnarray}
while the eigenstates of $\tilde H^\sigma_{ij}$ are denoted
$|\phi^\sigma_{S,m}(i,j)$ with energies $-3/4$ for the singlet
($S=0$, $m=0$) and $1/4$ for the triplets ($S=1$, $m=0,\pm1$).

\begin{figure}[hp]
\centerline{\psfig{figure=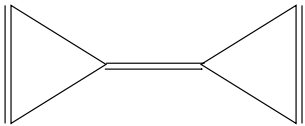,width=4.0cm,angle=0}}
\vspace{0.5cm}
\caption{Solution of the two-triangle problem. Singlets are represented as 
double lines.}
\label{fig2}
\end{figure}

It is useful to start with two
triangles coupled by a single link. In the spirit of the Majumdar-Ghosh solution
of the $J_1-J_2$ chain\cite{majumdar}, the groundstate wavefunction in terms of 
the spins 
$\vec S$ is obtained as the product
of three singlets involving respectively the link between the triangles and the
remaining two spins on each triangle (see Fig. 2), 
and its energy is $-6J/4-3J'/4$.
In that simple case, the effective Hamiltonian $\tilde H_{ij}$ is the 
product of a spin part $\tilde H^\sigma_{ij}$ and a chirality
part $\tilde H^\tau_{ij}$,
and its eigenvalues are the products of the 
eigenvalues of  $\tilde H^\sigma_{ij}$ 
with those of $\tilde H^\tau_{ij}$. The groundstate energy of $\tilde H_{ij}$ 
is thus $-3J'/4$: The exact
groundstate energy is recovered, as it should since it depends linearly on $J'$.

In the general case of Eq.(\ref{effective}), the Hamiltonian is not 
the product of a spin part and a chirality part, and its solution is in
principle as difficult as the original problem. However, we note that the
Hamiltonian of Eq.(\ref{effective}) is formally similar to the Kugel--Khomskii  
model that
was introduced in the context of orbitally degenerate magnets\cite{kugel}, 
and a mean-field decoupling of the 
spin and orbital degrees of freedom is known to give an accurate picture of the
physics when the assymmetry between spin and orbital degrees of freedom is
strong enough\cite{castellani}. In our case, the assymmetry between spin and
chiral variables in Eq.(\ref{effective}) is clearly very strong, and such a 
decoupling amounts 
to the replacement of $\tilde H$
with the mean-field Hamiltonian:
\begin{equation}
H_{MF} = \sum_{<i,j>} (a^\tau_{ij} \tilde H^\sigma_{ij} + a^\sigma_{ij} 
\tilde H^\tau_{ij} - a^\sigma_{ij} a^\tau_{ij} )
\label{mean-field}
\end{equation}
where the parameters $a^\tau_{ij}\equiv <\tilde H^\tau_{ij}>$ and 
$a^\sigma_{ij}\equiv <\tilde H^\sigma_{ij}>$ have to be determined
self-consistently. Note that this mean-field problem is still very complicated
{\it a priori} since it involves S=1/2 Heisenberg like models on a
triangular lattice.

Remarkably enough, the low-energy solutions of 
that problem can be determined
analytically. Let us concentrate for the moment on clusters with an even number
of sites and with periodic boundary conditions, and let us consider a 
dimer covering of the triangular 
lattice by nearest-neighbor dimers. Denoting by $D$ the set of nearest-neighbor 
pairs that enter this covering, we can construct a wavefunction $|\Phi_0(D)>$
in the following way:
\begin{equation}
|\Phi_0(D)> = \prod_{<i,j>\in D} |\phi^\tau_{1}(i,j)> \otimes 
|\phi^\sigma_{0,0}(i,j)>
\label{phi0}
\end{equation}
Clearly $|\Phi_0(D)>$ will be a solution of the problem if $a^\tau_{ij}=
a^\sigma_{ij}=0$ as soon as $<i,j>\notin D$. This turns out to be true thanks 
to the
following properties
\begin{eqnarray}
<\phi^\tau_{1}(i,j)\phi^\tau_{1}(k,l)|\tilde H^\tau_{jk}|
\phi^\tau_{1}(i,j)\phi^\tau_{1}(k,l)> & = & 0 \nonumber \\
<\phi^\sigma_{0,0}(i,j)\phi^\sigma_{0,0}(k,l)|\tilde H^\sigma_{jk}|
\phi^\sigma_{0,0}(i,j)\phi^\sigma_{0,0}(k,l)> & = & 0 
\end{eqnarray}
which can be easily checked directly with the expressions of the wavefunctions.
So $|\Phi_0(D)>$ is a solution characterized  by 
$a^\tau_{ij}=9, a^\sigma_{ij}=-3/4$ 
if $<i,j>\in D$ 
and 0 otherwise, its energy is given by $E_0(S=0)=-(3J'/8)N_t$, and it is a
singlet\cite{note2}. 
To prove that is minimizes the energy of Eq.(\ref{mean-field}), we have solved
the mean-field problem  numerically on small clusters with 
up to $3\times 4$ sites. This has been done by iteration starting from random
values of $a^\sigma_{ij}$ and $a^\tau_{ij}$, and we found that the lowest 
energy is always equal to $-(3J'/8)N_t$.
In fact, solutions like $\Phi_0(D)$ exist most of the time for such mean-field 
Hamiltonians, but they are usually significantly higher in energy than 
uniform solutions.
In the present case however, 
other solutions involving $|\phi^\tau_{2}(i,j)>$ 
are very bad energetically because $E_2=E_1/9$. 
Note that the wavefunction $|\phi^\tau_{1}(i,j)>$ corresponds neither to
ferromagnetic nor antiferromagnetic ordering of the chiral variables since it is
a linear combination of the four basis states $|RR>$, $|RL>$, $|LR>$ and $|LL>$.
So there is no chiral ordering of the type discussed by Baskaran\cite{baskaran} 
for the triangular lattice even locally. 

Similarly, triplet solutions can be constructed for a given dimer covering of 
the
triangular lattice. Choose two neighbouring sites $(i_0,j_0)$, and consider a
dimer covering $D(i_0,j_0)$ of the remaining sites. The wavefunction with 
lowest energy is of the form
\begin{equation}
|\Phi_0(D(i_0,j_0))>
(|\phi^\tau_{n}(i_0,j_0)> \otimes 
|\phi^\sigma_{1,m}(i_0,j_0)>)
\label{S=1}
\end{equation}
In this expression, $m=0,\pm1$ and $n$ can take the values 3 or 4\cite{note3}. 
The energy
of this state is $E_0(S=1)=E_0(S=0)+(2/3)J'$. Again it 
was
checked numerically on a $3\times 4$ cluster that this is indeed the lowest 
energy in the triplet sector. So this mean-field approach predicts that there 
is a singlet-triplet energy gap $\Delta$ equal to $(2/3)J'$.

Another class of low-lying states exist in the triplet sector. They can be
constructed in the following way: Choose two sites $(k,l)$ that are 
{\it not} nearest neighbours, denote by $D(k,l)$ a dimer covering of the
remaining sites, and consider the wavefunction
\begin{equation}
|\Phi_0(D(k,l))> |\sigma_k \tau_k> |\sigma_l \tau_l>
\end{equation}
where $|\sigma_k \tau_k>$ (resp. $|\sigma_l \tau_l>$) can be any configuration 
at site $k$ (resp. $l$). Then similar arguments show that this is a solution
with energy $E_1(S=1)=E_0(S=1)+J'/12$. Each unpaired site
correspond to a $S=1/2$ excitation and can be interpreted as a
spinon. This mean-field approach predicts that spinons form triplet 
bound states
on neighbouring sites with a binding energy 
equal to $J'/12$. In fact, if we consider a cluster with an odd number of sites,
the groundstate can be shown to consist of one unpaired site - i.e. one spinon -
times $|\Phi_0(D)>$, where $D$ is a dimer covering of the remaining sites.

Now let us turn to the very interesting question of the groundstate degeneracy.
The energy $E_0(S=0)$ does not depend on the particular dimer covering
$D$ used to construct the wavefunction $|\Phi_0(D)>$. So, for a given cluster, 
the degeneracy is controlled by the number of dimer coverings. 
This number can be calculated using standard techniques\cite{fisher,samuel}. 
For the triangular lattice, we found that it increases with the number of sites
$N_t$ like $\alpha^{N_t}$ with
$\ln \alpha=\frac {1} {16\pi^2} \int_0^{2\pi}  \int_0^{2\pi}  \ln (4+4 \sin x
\sin y + 4 \sin^2 y) dx dy$.
A numerical integration yields $\ln \alpha = 0.4286$, or $\alpha=1.5351$. 
In terms of the
original Kagome lattice, this corresponds to a degeneracy 
that increases like $(\alpha^{1/3})^N = 1.1536^N$
since $N=3N_t$. 

Next let us try to assess the validity of this mean-field approach. One
argument 
in favour of this kind of approach is that it is qualitatively correct in the
case of three coupled chains with periodic boundary conditions: Schulz's 
analysis based on Renormalization Group argument predicts a dimerized, twofold
degenerate
groundstate with gapped spin excitations\cite{schulz,kawano}, and this is 
exactly the 
physics of the
mean-field solution adapted to that case\cite{mila}.
To check its validity in cases of higher groundstate degeneracy,
we have performed numerical
simulations of the effective Hamiltonian of Eq.(\ref{effective}) for finite
systems of coupled triangles. For the modified Kagome problem 
this turns out to be very difficult because the first cluster where dimer
coverings can be done properly has $3 \times 4$ sites. To determine the low
lying states is then a numerical task comparable to the Kagome problem itself 
for 36 sites, and this is beyond the scope of the present paper. So we have
compared mean-field and exact results for similar but different systems, namely
for rings of $n$ triangles with diagonal bonds up to $n=6$. 
For $n=2$, the mean-field description is of course 
exact. For $n=4$ and $6$, the structure of the low-energy spectrum is correctly
given by the mean-field approach. For instance, for $n=4$, the mean-field
approach predicts that the groundstate is threefold degenerate, and the exact 
results 
give 3 low-lying singlets at $-1.72053 J'$ (non degenerate) and $-1.54077 J'$ 
(two-fold degenerate), the next state being a triplet located at $-1.13942 J'$.
For $n=6$, the groundstate degeneracy is 6 at the mean-field level, and the
exact results also give 6 low-lying singlets. The details will be
given elsewhere, but the answer is clear: The mean-field approach is
qualitatively correct
in predicting the number of low-lying
singlets. To go beyond mean-field is expected to partially lift the degeneracy
within the groundstate manifold but not to change the number of low-lying
singlets. 

Finally, let us come to the most important question: What can we learn from this
approach concerning the regular (non-dimerized) Kagome lattice? 
First of all, let us translate our results in terms of the original spins $\vec
S$. 
We know from the analysis of the two triangle
problem that the basic brick of our mean-field wavefunction, namely a two-site 
wavefunction of
the type  $|\phi^\tau_{1}(i,j)> \otimes |\phi^\sigma_{0,0}(i,j)>$, 
corresponds to a dimer mapping of the two triangle problem 
(see Figure 2). So, the wavefunctions of Eq.(\ref{phi0}) correspond to a 
certain subset of the dimerized wavefunctions of the Kagome lattice 
used by Zeng and Elser\cite{zeng2}. 
Next, let us compare the physical properties of this mean-field solution with
the numerical results obtained on the Kagome lattice. As we noticed in the 
introduction,
the number of dimerized states increases like $1.26^N$, i.e much too fast. 
However, the number of states selected with our criterion increases roughly 
like 
$1.15^N$, 
in agreement with the numerical results for even clusters\cite{bernu}. 
This is probably the
most interesting result of the present approach: It provides 
a simple but
nevertheless quantitative explanation of the very numerous low-lying singlet 
states of the S=1/2 Kagome antiferromagnet. Their number scales with the number
of dimer coverings of the underlying triangular lattice of triangles.
The fact that one can choose triangles pointing upwards or downwards to build
the wave-function means that the actual low lying states can have domains with
different orientations of the triangles. Elementary energy considerations
suggest 
that these domains are large however, so that they can only marginally 
contribute to the increase of the number of singlet states with the size of the
system.
Interestingly enough we can also explain
the apparent discrepancy between odd and even clusters in the results of
Ref.\cite{bernu}. For odd clusters, there is an unpaired site, and according to
the present theory, the
degeneracy is expected to scale like $ N \times 1.15^N$. Omitting this prefactor
in the fit leads to an overestimate of $\alpha$ (see Figure 3). 

\begin{figure}[hp]
\vspace{-1.cm}
\centerline{\psfig{figure=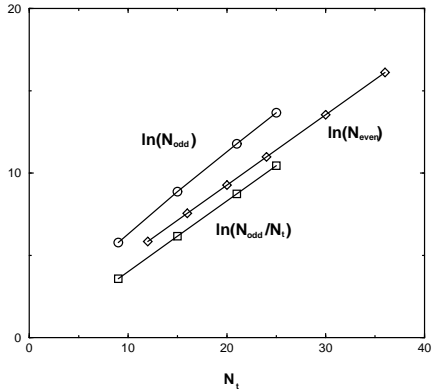,width=7.0cm,angle=-90}}
\vspace{0.5cm}
\caption{Logarithm of the number of dimer coverings of the triangular lattice as
a function of the number of sites $N_t$ for small systems with periodic boundary
conditions. The slope of $\ln (N_{odd})$ is 1.18, while the slope of both
$\ln (N_{even})$ and $\ln (N_{odd}/N_t)$ is 1.15.}
\label{fig3}
\end{figure}

To summarize, we have studied a dimerized version 
of the S=1/2 Kagome antiferromagnet by performing a mean-field analysis 
of the effective
hamiltonian that describes its low-energy sector. This approach leads to a 
transparent picture of the low energy properties which turns out to bear 
remarkable similarities to those reported for the regular Kagome
lattice\cite{bernu}, e.g. the number of low-lying singlets or the presence 
of a singlet-triplet gap. Besides, these results lead to natural subspaces of 
dimer wavefunctions to describe the low energy singlet and triplet sectors of 
the regular Kagome model. A variational study of this model using these 
wave-functions is in progress to make more precise statements about the accuracy
of this description of the low energy sector of the regular Kagome
antiferromagnet. It is expected to give useful
information concerning the structure of the low-lying singlet sector, for
instance the low temperature specific heat, and to allow one to make
more precise predictions concerning the persistence of the singlet-triplet gap
and the binding of spinons in the thermodynamic limit.

I would like to thank Claire Lhuillier, Maurice Rice and Heinz Schulz 
for very interesting discussions. I would also like to acknowledge the 
hospitality of the ETH Z\"urich where this project started. The 
numerical simulations were performed on the Cray supercomputers 
of the IDRIS (Orsay, France).


\begin{references}
\bibitem{note1}  For 36 sites, this subspace is already larger
than the actual low-lying singlet subspace by a factor 25!
\bibitem{note2} Note that although this mean-field energy has already been 
derived in Ref.\cite{subrahmanyam}, no attempt was made to discuss 
the degeneracy 
of the singlet manifold or to study the triplet sector, 
which are clearly the most important results of the present work.
\bibitem{note3} The argument is slightly more involved though 
than for the singlet groundstates
since the order parameters $a_{ij}^\tau$ do not necessarily vanish. 
It will be
reproduced elsewhere.
\end{references}
\end{document}